\documentstyle[psfig]{aa}

\def\BD#1#2{BD{$#1$}$^\circ${#2}}
\def\bd#1#2{{$#1$}$^\circ${#2}}

\begin{document}
   \thesaurus{06   
             (08.01.1    
              08.16.4    
              08.02.5    
              08.05.3    
              08.12.1 )} 

   \title{The Henize sample of S stars}

   \subtitle{I. The technetium dichotomy\thanks{
Based on observations carried out at the 
European Southern Observatory (ESO, La Silla, Chile)}
 }
 
   \author{S. Van Eck 
      \and A. Jorissen
      \thanks{Research Associate, National Fund for 
              Scientific Research (FNRS), Belgium}}

   \offprints{S. Van Eck}

   \institute{    Institut d'Astronomie et d'Astrophysique (IAA),
                  Universit\'e Libre de Bruxelles, C.P.226,
                  Boulevard du Triomphe,
                  B-1050 Bruxelles,
                  Belgium\\
                  email: svaneck@astro.ulb.ac.be, ajorisse@astro.ulb.ac.be
}

   \date{Received date; accepted date}

   \maketitle

   \begin{abstract}
This paper is the first one in a series investigating the properties
of the S stars belonging to the Henize sample (205 S stars
with $\delta<-25^\circ$ and $R<10.5$) in order to derive the respective
properties (like galactic distribution and relative frequencies)
of intrinsic (i.e. genuine asymptotic giant branch) 
S stars and extrinsic (i.e.
post mass-transfer binary) S stars.
High-resolution (R=30\,000 to 60\,000) spectra covering 
the range $\lambda\lambda4230-4270$\AA~ 
have been obtained for 76 S stars, 8 M stars and 2 symbiotic stars.

The $\lambda4262$\AA~ and $\lambda4238$\AA~ blends involving a \ion{Tc}{i}
line were analysed separately and yield consistent conclusions 
regarding the presence or absence of technetium.
Only one `transition' case (Hen 140 = HD 120179, a star where
only weak lines of technetium are detectable) is found in
our sample. 
A resolution greater than R $=30\,000$ is clearly required in order
to derive unambiguous conclusions concerning the presence or absence of
technetium.
The Tc/no Tc dichotomy will be correlated
with radial velocity and photometric data in a forthcoming paper.

   \end{abstract}

\section{Introduction}

S stars have been identified as a class of peculiar red giants by 
Merrill (\cite{Merrill22}). 
Basically, the S stars emerge as a sequence parallel
to the M stars as far as temperature is concerned, but with enhanced 
abundances of s-process elements. The chief observational
difference between M and S spectra is the presence of ZrO bands
in the latter.
The s-process overabundances in S stars are explained in the
framework of the Thermally Pulsing Asymptotic Giant Branch (TPAGB) evolution.
Thermal instabilities (called {\it thermal pulses}) that affect 
the helium-burning shell of these stars have two important consequences:
they provide the proper environment for the nucleosynthesis 
of s-process elements, and they trigger an envelope response
(the {\it third dredge-up}) that allows the s-process elements and carbon 
to be brought to the surface [see e.g. Mowlavi (\cite{Mowlavi}) for a review].

Technetium is an s-process element with no stable isotope that was first 
identified in the spectra of some M and S stars by Merrill (\cite{Merrill52}). 
$^{99}$Tc, with a laboratory half-life of $t_{1/2} = 2.13\times 10^5$ yr,
is the only technetium isotope produced by the s-process.
The high temperatures encountered during
thermal pulses strongly shorten the effective half-life of $^{99}$Tc
($t_{1/2}\sim 1$ yr at $\sim 3\times 10^8$ K, Cosner et al. \cite{Cosner})
but the large neutron densities at these high temperatures
more than compensate the reduction of $^{99}$Tc life-time 
(Mathews et al. \cite{Mathews}) and enable a substantial 
technetium production.
Third dredge-up episodes then carry technetium to the envelope, where it decays
steadily at its terrestrial rate of $t_{1/2}=2.13 \times 10^5$ yr.
Starting from an abundance corresponding to the maximum observed one,
technetium should remain detectable during $1.0-1.5\times 10^{6}$ yr
(Smith \& Lambert \cite{Smith}).
If the dredge-up of heavy elements 
occurs after each thermal pulse 
(occurring every $\sim 1-3\times 10^{5}$ yr), 
virtually all s-process enriched TPAGB stars should exhibit technetium lines.

However, Straniero et al. (\cite{Straniero}) advocated that the s-process 
nucleosynthesis mainly occurs during the interpulse.
When technetium is engulfed in the subsequent thermal pulse,
it will decay at a fast rate because of the high temperature,
and will not be replenished if there is no neutron source
operating within the pulse itself.
The conclusion that s-process enriched TPAGB stars should necessarily
exhibit technetium would then be challenged. 
Nevertheless, all the S stars identified as TPAGB stars 
by Van Eck et al. (\cite{Van Eck98}) 
thanks to the HIPPARCOS parallaxes
turned out to be Tc-rich.
A survey of technetium in a large sample of S stars like the
Henize sample may be expected to provide further constraints
on the s-process environment in AGB stars 
(e.g. interpulse s-process versus thermal-pulse s-process,
thermal-pulse duration and temperature versus $^{99}$Tc half-life).

Not all S stars exhibit Tc lines though 
(Little-Marenin \& Little \cite{Little-Marenin}, 
Little et al. \cite{Little}),
but technetium-poor stars (also called {\it extrinsic}, 
as opposed to technetium-rich, {\it intrinsic} S stars)
are currently believed to emerge from a totally different evolutionary history:
because they are members of binary systems
(Brown et al. \cite{Brown}, Jorissen et al. \cite{Jorissen93}, 
Johnson et al. \cite{Johnson}, Jorissen et al. \cite{Jorissen}), 
they rather owe their chemical peculiarities to the accretion of 
s-process-rich matter from their companion (formerly a TPAGB star,
now an undetected white dwarf). 
They are technetium-poor, because enough time has elapsed for the technetium
to decay since the mass transfer event.
The $^{99}$Tc half-life is indeed much shorter than any stellar
evolutionary timescale (but the TPAGB).
Such a polluted giant star will be classified either as a G or K giant
with enhanced heavy elements (i.e. as a barium star) or, if it has
cooled enough for the ZrO molecular bands to appear, as a technetium-poor
S star.

Besides technetium detection, several spectroscopic criteria
(of various efficiencies) aiming at distinguishing
extrinsic from intrinsic S stars have been
mentioned in the literature [e.g. oxygen isotopic ratio (Smith
\& Lambert \cite{Smith90}), presence of the \ion{He}{I} $\lambda 10830$
line (Brown et al. \cite{Brown}), zirconium isotopic
ratio (Busso et al. \cite{Busso})], but technetium detection appears to be,
by far, the most secure and tractable way to unmask extrinsic S stars.

This unmasking operation is crucial when deriving fundamental 
stellar quantities such as the third dredge-up luminosity threshold
(commonly measured as the minimum luminosity of S stars).
Evolutionary timescales of TPAGB stars can be strongly in error
if the considered star samples are polluted by non-AGB, 
mass-transfer S stars.

We therefore decided to study a large and properly defined sample of S stars
in order to disentangle the two sub-families and to study their
respective characteristics. 
The Henize sample of S stars 
(Henize \cite{Henize}, as listed in Stephenson \cite{Stephenson}) 
consists of 205 S stars
south of $\delta=-25^\circ$ and brighter than $R=10.5$.
Radial velocity data, low- and high-resolution spectroscopy,
as well as Geneva photometry have been collected over several years.
The present paper deals with high-resolution technetium spectra
for 72 Henize stars. Some additional
K, M and symbiotic stars data are also presented.
Results concerning binarity and photometry, as well as the global analysis
of the different data sets, are postponed to a forthcoming
paper. 

\section{Observations and reduction}
\subsection{Instrumental set-up}
\label{Sect:setting}

The high-resolution spectra used in the present study were  
obtained during several runs (1991-1998) at the European 
Southern Observatory, with the Coud\'e Echelle Spectrometer 
(CES) fed by the 1.4m Coud\'e Auxiliary Telescope (CAT).
The 1991-1993 runs were performed with the short camera (f/1.8)
and CCD \#9 (RCA SID 503 thinned, backside illuminated, 
$1024 \times 640$ pixels of $15~\mu$m), 
whereas the long camera (f/4.7) and CCD \#38 
(Loral/Lesser thinned, backside illuminated, UV flooded, $2688 \times 512$ 
pixels of $15~\mu$m) were used during the 1997-1998 runs.
Details on these configurations can be found in Lindgren \& Gilliotte 
(\cite{Lindgren}) and Kaper \& Pasquini (\cite{Kaper}).
The resolution ran\-ges from 0.14\AA~ (R=30\,000) 
to 0.07\AA~ (R=60\,000) for a 
central wavelength of 4250\AA. The spectra approximately cover
the wavelength range $\lambda\lambda$ 4230-4270\AA.

\subsection{Stellar samples}

The observed stars are a subset of the sample of 205 S stars
collected by Henize (\cite{Henize}) from his objective-prism survey 
(with a dispersion of 450\AA/mm at H$\alpha$) of ZrO stars
south of $\delta = -25^{\circ}$ and brighter than $R=10.5$.
Given the limitations on the CAT pointing and on the detectors sensitivity,
only stars with $\delta > -75^{\circ}$, $V<11$ and $B-V<2$
(translating into 70 objects) 
could be observed in a reasonable amount of time, 
i.e. less than 1h30 per star.
A few bright redder stars could also be observed 
(but see the discussion on SC stars in Sect.~\ref{Sect:SC}).
A sample of bright M stars with an excess at $60\mu$m [indicative of a
possibly detached dust shell; see Zijlstra et al. (\cite{Zijlstra})],
as well as the two symbiotic stars RW Hya and SY Mus,
and some radial-velocity standards have also been observed.
Three non-Henize S stars from our radial-velocity monitoring
(Udry et al. \cite{Udry}) have been included as well. 
The log of the observations, including the instrumental setting,
is given in Table~\ref{Tab1} and, for Henize stars, in Table~\ref{Tab2}.

\renewcommand{\baselinestretch}{0.8}
  \begin{table*}
  \caption[]{\label{Tab1} Observations log and results 
for non-Henize stars.\\
The first column identifies the star by its HD, HR or BD number.
The spectral type is quoted next.
The next two columns list the civil date of the observation (day month year) 
and the Julian date (JD - 2\,448\,000.5).
The column CCD provides the instrumental setting 
(9-sh: CCD \#9 $+$ short camera; 38-l: CCD \#38 $+$ long camera).
The next column indicates the spectral resolution.
The next six columns give, for the two considered technetium blends
(at 4262\AA~ and 4238\AA, respectively),
the (Doppler-shift corrected) central wavelength of the technetium blend
($\lambda$ - 4200\AA), 
the standard deviation on the Doppler shift ($\sigma$, in \AA)
and the signal-to-noise ratio (see text). The technetium-rich (y) or
technetium-poor (n) status is given next, while the last column contains 
additional remarks
}

  \small{
  \setlength{\tabcolsep}{1.2mm}
  \begin{tabular}{lllrrc|ccr|ccr|cl}
HD/BD   &type    & cdate&Jdate&CCD &    R&\multicolumn{3}{|c|}
{4262\AA~ blend}&\multicolumn{3}{|c|}{4238\AA~ blend} &Tc&Remarks \cr
\rule[-2mm]{0mm}{5mm}
        &        &      &      &    &     & $\lambda$ & $\sigma$ & 
S/N &  $\lambda$ & $\sigma$ & S/N &    & \cr \hline 
\multicolumn{14}{l}{\rule[-2mm]{0mm}{6mm}$\bullet$ M stars}\\
HR 4938 &M3.5III &160393&1063.3&9-sh&45000&62.104&0.012&  41&38.366&0.028&
  29&n &V789 Cen \cr             
HR 5064 &K5-M0III&150393&1062.3&9-sh&45000&62.097&0.010&  76&38.353&0.030&
  61&n & \cr                  
HR 5134 &M5III   &150393&1062.3&9-sh&45000&62.069&0.015&  70&38.356&0.037&
  53&n &V744 Cen \cr            
\multicolumn{14}{l}{\rule[-2mm]{0mm}{6mm}$\bullet$ M stars with 
60$\mu$m excess}\\
73341   &M3/M4III&300593&1138.0&9-sh&45000&62.103&0.011&  33&38.370&0.030&
  25&n &SAO 236108 \cr         
91094   &M1III   &300593&1138.0&9-sh&45000&62.099&0.014&  39&38.357&0.025&
  34&n &SAO 250981 \cr         
179199  &M2III   &290593&1137.4&9-sh&45000&62.097&0.017&  18&38.354&0.032&
  15&n &SAO 162305 \cr         
181620  &M2III   &290593&1137.4&9-sh&45000&62.109&0.016&  29&38.368&0.025&
  23&n &V4415 Sgr;SAO 211215 \cr
\multicolumn{14}{l}{\rule[-2mm]{0mm}{6mm}$\bullet$ Symbiotic stars}\\
100336  &M4.5    &160393&1063.2&9-sh&45000&62.094&0.020&  11&38.347&0.029&
   8&n &SY Mus  \cr    
117970  &M2      &150393&1062.2&9-sh&45000&62.115&0.011&  31&38.363&0.042&
  24&n &RW Hya  \cr    
\multicolumn{14}{l}{\rule[-2mm]{0mm}{6mm}$\bullet$ Non-Henize S stars}\\
1760    &M5-6Se  &300792& 834.4&9-sh&30000&62.230&0.020&  65&38.142&0.030&
  55&y &T Ceti (also M5-6Ib-II)\cr    
\bd{-21}{2601}&S3*3    &310197&2480.2&38-l&60000&62.097&0.007&  22&38.365&
0.024&  16&n & \cr
\bd{-08}{1900}&S4/6    &240298&2869.0&38-l&60000&62.239&0.029&  15&38.256&
0.035&  11&y & \cr          
\bd{+04}{4356}&S4*3    &290593&1137.3&9-sh&45000&62.210&0.015&  19&38.079&
0.033&  11&y & see note \cr        
\multicolumn{14}{l}{\rule[-2mm]{0mm}{6mm}$\bullet$ Radial velocity standards}\\
80170   &K5III   &240298&2869.1&38-l&60000&62.099&0.011&  84&38.356&0.018&
  66&n & \cr           
\rule[0mm]{0mm}{2mm}108903  &M3.5III &240298&2869.3&38-l&60000&62.104&
0.012& 233&38.372&0.016& 182&n &$\gamma$ Cru A 
\rule[-2mm]{0mm}{2mm} \cr
\hline
\rule[-2mm]{0mm}{2mm} \cr
\end {tabular}
Note: The S star \BD{+04}{4356} = GCGSS 1193 = SAO 125493 = IRAS 20062+0451
has been erroneously associated by MacConnell (\cite{MacConnell}) 
with the nearby (non-S) star \BD{+04}{4354}. The coordinates
in the original paper are nevertheless correct.
Since then, this error has propagated in the literature
(Stephenson \cite{Stephenson}, Jorissen et al. \cite{Jorissen93}, 
Chen et al. \cite{Chen}, Jorissen et al. \cite{Jorissen}, 
Udry et al. \cite{Udry}), 
although in all these papers the measured star was indeed
the S star \BD{+04}{4356}.
}
\end{table*}

  \begin{table*}
  \caption[]{\label{Tab2} Same as Table~\ref{Tab1} but for Henize S stars.
The Henize and GCGSS  numbers (Stephenson \cite{Stephenson}) are listed in 
the first two columns. Spectral types are from Stephenson (\cite{Stephenson})
}
  \small{
  \setlength{\tabcolsep}{1.2mm}
  \begin{tabular}{rrlllrlc|ccr|ccr|cl}
Hen&GC-&HD/DM &type & cdate& Jdate&CCD &    R&\multicolumn{3}{|c|}
{4262\AA~ blend}&
\multicolumn{3}{|c|}{4238\AA~ blend} &Tc&Remarks \cr
\rule[-2mm]{0mm}{5mm}
   & GSS &         &     &      &      &    &     & $\lambda$ & $\sigma$ & 
S/N &  $\lambda$ & $\sigma$ & S/N &    & \cr \hline \rule[-1mm]{0mm}{5mm}
  1&    3&310      &S3,1 &210991& 521.2&9-sh&30000&62.107&0.008&  36&
38.375&0.013&  34&n &\cr                
  2&   39&9810     &S2,1 &210991& 521.3&9-sh&30000&62.094&0.015&  21&
38.372&0.014&  18&n &\cr                
  3&  104&29704    &S:   &210991& 521.3&9-sh&30000&62.088&0.019&  26&
38.396&0.016&  23&n &\cr                
  5&  139&         &S3,3 &010297&2481.0&38-l&60000&62.089&0.007&  18&
38.394&0.010&  12&n &\cr                
  6&  141&         &S5,2 &300197&2479.1&38-l&60000&62.091&0.008&  22&
38.379&0.015&  11&n &\cr                
  7&  178&40706    &S2,1 &150393&1062.0&9-sh&45000&62.126&0.022&  22&
38.367&0.025&  20&n &\cr                
  8&  202&\bd{-39}{2449} &S5,6 &010297&2481.1&38-l&60000&62.227&0.020&
  18&38.194&0.042&  14&y &\cr                
  9&  204&\bd{-60}{1381} &S3,3 &160393&1063.0&9-sh&45000&62.099&0.008&
  16&38.349&0.024&  13&n &\cr                
 14&  242&\bd{-34}{3019} &S2,5 &160393&1063.1&9-sh&45000&62.104&0.004&
  13&38.363&0.023&  10&n &\cr                
 16&  248&         &S:   &310197&2480.2&38-l&60000&62.258&0.019&  13&
38.172&0.014&   7&y &\cr                
 18&  294&\bd{-28}{3719} &S6,8e&150393&1062.0&9-sh&45000&62.072&0.032&
  13&38.378&0.033&  10&n &\cr                
 19&  328&         &     &310197&2480.1&38-l&60000&62.229&0.015&  16&
38.233&0.021&  12&y &\cr                
 20&  342&\bd{-45}{3132} &     &310197&2480.0&38-l&60000&62.230&0.016&
  27&38.120&0.017&  16&y &\cr                
 28&  390&62340    &S4,4 &010297&2481.1&38-l&60000&62.079&0.016&  36&
38.381&0.011&  26&n &\cr                
 31&  434&65152    &S1,1 &150393&1062.1&9-sh&45000&62.093&0.009&  31&
38.369&0.026&  27&n &\cr                
 34&  446&         &S7,2 &010297&2481.1&38-l&60000&62.254&0.022&  39&
38.193&0.013&  30&y &X Vol \cr          
 35&  447&\bd{-71}{435}  &S1,1 &300197&2479.1&38-l&60000&62.078&0.010&
  19&38.375&0.026&  11&n & \cr
 36&  448&\bd{-31}{5393} &S3,1 &150393&1062.1&9-sh&45000&62.222&0.011&
  22&38.141&0.028&  17&y &\cr                
 37&  456&\bd{-41}{3702} &S4,2 &010297&2481.2&38-l&60000&62.260&0.004&
  19&38.090&0.017&  12&y &\cr                
 39&  461&\bd{-65}{601}  &S6,2 &010297&2481.2&38-l&60000&62.238&0.004&
  31&38.118&0.014&  21&y &\cr                
 41&  474&\bd{-27}{5131} &S4,2 &010297&2481.2&38-l&60000&62.219&0.027&
  31&38.210&0.019&  23&y &\cr                
 43&  487&\bd{-26}{5801} &S4,4 &240298&2869.1&38-l&60000&62.099&0.012&
  15&38.348&0.026&  10&n &\cr                
 45&  490&\bd{-32}{5117} &     &010297&2481.3&38-l&60000&62.212&0.015&
  25&38.147&0.019&  18&y &\cr                
 57&  559&         &S4,2 &310197&2480.2&38-l&60000&62.221&0.021&  21&
38.145&0.019&  14&y &\cr                
 63&  588&\bd{-33}{5772} &S4,1 &300197&2479.2&38-l&60000&62.103&0.013&
  15&38.364&0.017&  10&n &\cr                
 64&  591&\bd{-28}{6970} &S7/5e&010297&2481.3&38-l&60000&62.252&0.010&
  17&38.181&0.031&   7&y &\cr                
 66&  593&\bd{-33}{5803} &S5,2 &310197&2480.3&38-l&60000&62.226&0.029&
  18&38.138&0.013&  12&y &\cr                
 79&  653&         &S5,2 &240298&2869.2&38-l&60000&62.097&0.011&  19&
38.388&0.023&  14&n &\cr                
 80&  656&         &S5,6 &150393&1062.2&9-sh&45000&62.254&0.017&  15&
38.138&0.022&  12&y &KN Car \cr         
 88&  667&\bd{-30}{8296} &S5,2 &300197&2479.3&38-l&60000&62.220&0.012&
  15&38.122&0.008&  10&y &\cr                
 89&  668&         &S3,1 &300197&2479.3&38-l&60000&62.223&0.018&  18&
38.091&0.012&   9&y &\cr                
 90&  672&\bd{-54}{3378} &S5,6 &240298&2869.2&38-l&60000&62.083&0.008&
  22&38.372&0.021&  16&n &\cr                
 95&  693&         &S4,2 &300593&1138.0&9-sh&45000&62.229&0.016&  12&
38.138&0.020&   8&y &\cr                
 97&  696&         &S5,2 &290593&1137.1&9-sh&45000&62.224&0.006&  21&
38.148&0.033&  16&y &HP Vel \cr         
101&  704&         &S5,4 &240298&2869.3&38-l&60000&62.247&0.037&  21&
38.165&0.032&  15&y &Z Ant \cr          
104&  714&95013    &S5,4 &300593&1138.1&9-sh&45000&62.208&0.015&  10&
38.176&0.008&   6&y &\cr                
108&  720&95875    &S3,3 &300197&2479.3&38-l&60000&62.087&0.014&  34&
38.368&0.017&  21&n &\cr                
119&  778&104361   &S3,3 &010297&2481.4&38-l&60000&62.084&0.010&  27&
38.377&0.015&  16&n &\cr                
121&  792&\bd{-27}{8661} &S4,6e&300593&1138.1&9-sh&30000&62.098&0.016&
  10&38.374&0.035&   8&n &\cr                
123&  795&\bd{-47}{7642} &S4,2 &160393&1063.3&9-sh&45000&62.096&0.013&
  20&38.358&0.027&  14&n &CSV101280 \cr      
126&  802&         &S4,2 &240298&2869.4&38-l&60000&62.100&0.018&  17&
38.368&0.022&  12&n &\cr                
129&  808&\bd{-46}{8238} &S4,4 &290593&1137.1&9-sh&45000&62.073&0.018&
  14&38.384&0.034&  11&n &\cr                
132&  813&\bd{-72}{869}  &S4,6 &300593&1138.2&9-sh&30000&62.081&0.011&
  12&38.355&0.038&   9&n &\cr                
133&  814&114586   &S5,4 &240298&2869.3&38-l&60000&62.073&0.008&  20&
38.367&0.021&  16&n &\cr                
137&  824&\bd{-50}{7894} &     &160393&1063.3&9-sh&45000&62.088&0.013&
  22&38.366&0.028&  14&n &\cr                
138&  826&118685   &S6,2 &150393&1062.3&9-sh&45000&62.092&0.008&  32&
38.367&0.027&  25&n &-71 963 \cr        
140&  832&120179   &S3,1 &310197&2480.3&38-l&60000&  --  &  -- &  36&
  --  & --  &  27&y &see text\cr                
140&  832&120179   &S3,1 &300593&1138.2&9-sh&30000&  --  &  -- &  26&
  --  & --  &  20&y &see text\cr                
141&  834&120460   &S8,5 &230597&2592.2&38-l&60000&62.250&0.044&  19&
38.181&0.018&  12&y &VX Cen \cr         
143&  839&122434   &S3,1 &010297&2481.4&38-l&60000&62.092&0.013&  27&
38.372&0.019&  19&n &-41 8409 \cr       
147&  858&\bd{-25}{10393}&S3,3 &290593&1137.2&9-sh&45000&62.084&0.024&
   9&38.356&0.015&   7&n &\cr                
149&  864&130859   &S4,2 &150393&1062.4&9-sh&45000&62.102&0.007&  13&
38.370&0.023&  10&n &\cr                
150&  867&131217   &S6,2 &010892& 836.0&9-sh&30000&62.093&0.012&   6&
38.376&0.037&   5&n &\cr                
162&  927&         &S5,2 &300593&1138.3&9-sh&30000&62.239&0.016&  18&
38.181&0.039&  14&y &\cr                
173&  962&         &S4,2 &300593&1138.3&9-sh&30000&62.096&0.015&   9&
38.336&0.044&   7&n &\cr                
175&  974&156957   &S6/3+&310792& 835.1&9-sh&30000&62.243&0.032&  21&
38.142&0.027&  16&y &V635 Sco \cr       
177&  977&\bd{-32}{12687}&S:   &230597&2592.4&38-l&60000&62.232&0.032&
  24&38.107&0.027&  12&y & \cr 
178&  978&157335   &S5,4 &300593&1138.4&9-sh&30000&62.272&0.020&  10&
38.118&0.049&   6&y &V521 Oph \cr       
179&  994&160379   &S5,2 &150393&1062.4&9-sh&45000&62.090&0.012&  17&
38.369&0.025&  13&n &\cr                
182& 1010&163896   &S4,2 &310792& 835.2&9-sh&30000&62.082&0.008&  25&
38.376&0.027&  22&n &V745 Sgr \cr       
183& 1014&164392   &     &310792& 835.1&9-sh&30000&62.089&0.013&  26&
38.369&0.029&  22&n &\cr                
186& 1023&165774   &S4,6 &210991& 521.0&9-sh&30000&62.090&0.011&  30&
38.375&0.012&  26&n &\cr                
187& 1025&165843   &S2,1 &300593&1138.4&9-sh&30000&62.084&0.014&  20&
38.388&0.038&  14&n &\cr                
191& 1056&171100   &S5,4 &310792& 835.2&9-sh&30000&62.244&0.021&  19&
38.166&0.035&  15&y &V3574 Sgr \cr       
193& 1074&\bd{-23}{14695}&S4,2 &300792& 834.2&9-sh&30000&62.106&0.006&
   9&38.344&0.018&   8&n &\cr                
197& 1195&191630   &S4,4 &210991& 521.0&9-sh&30000&62.216&0.014&  50&
38.147&0.011&  42&y &\cr                
199& 1212&         &     &310792& 835.3&9-sh&30000&62.094&0.013&  14&
38.363&0.021&  12&n &\cr                
201& 1275&\bd{-26}{15676}&S3,3 &300792& 834.3&9-sh&30000&62.098&0.013&
  15&38.366&0.045&  13&n &\cr                
202& 1294&         &S5,7:&210991& 521.1&9-sh&30000&62.270&0.034&  27&
38.191&0.011&  17&y &$\pi^1$ Gru \cr         
203& 1295&         &S4,4 &300792& 834.3&9-sh&30000&62.067&0.017&  12&
38.358&0.033&  12&n &\cr                
204& 1303&         &S6,6 &310792& 835.3&9-sh&30000&62.116&0.016&  19&
38.368&0.023&  17&n &CSV103101 
\rule[-2mm]{0mm}{2mm} \cr
\hline
   \end{tabular}
}
\end{table*}
\renewcommand{\baselinestretch}{1.}

\subsection{Data reduction and S/N ratio}

The CCD frames were corrected for the electronic offset (bias)
and for the relative pixel-to-pixel response variations (flat-field).
Wavelength calibration was performed from thorium lamp spectra
taken several times per night. An optimal extraction of the
spectra was performed according to the method of Horne (\cite{Horne}). 
The whole reduction sequence was performed within
the `long' context of the MIDAS software package.

The signal-to-noise (S/N) ratio was estimated for each spectrum
in the following way: three S/N values were computed
for the three best exposed CCD lines (along the dispersion axis), 
in the neighborhood of the spectral region of interest 
(either 4262\AA~ or 4238\AA). These three
S/N values were then combined according to Eq. 17 of Newberry 
(\cite{Newberry}). 
When the exposure time on a given star has been split in two
(in order to reduce cosmics detrimental effect), the final S/N ratio
was computed using Eq. 18 of Newberry (\cite{Newberry}).
The degradation of the S/N ratio due to flat-field correction
has not been taken into account, since flat-fields have little
degrading effect for the low S/N values under consideration. 
The S/N ratio values are listed in Tables~\ref{Tab1} and ~\ref{Tab2} 
for each target star.
Because of the CCD spectral response, the S/N ratio near 4238\AA~ 
is systematically lower than the one near 4262\AA.

\section{Analysis}
\subsection{Fit of the technetium blends}
\label{Sect:Analysis}

\begin{figure}
   \begin{center}
   \leavevmode
   \centerline{\psfig{file=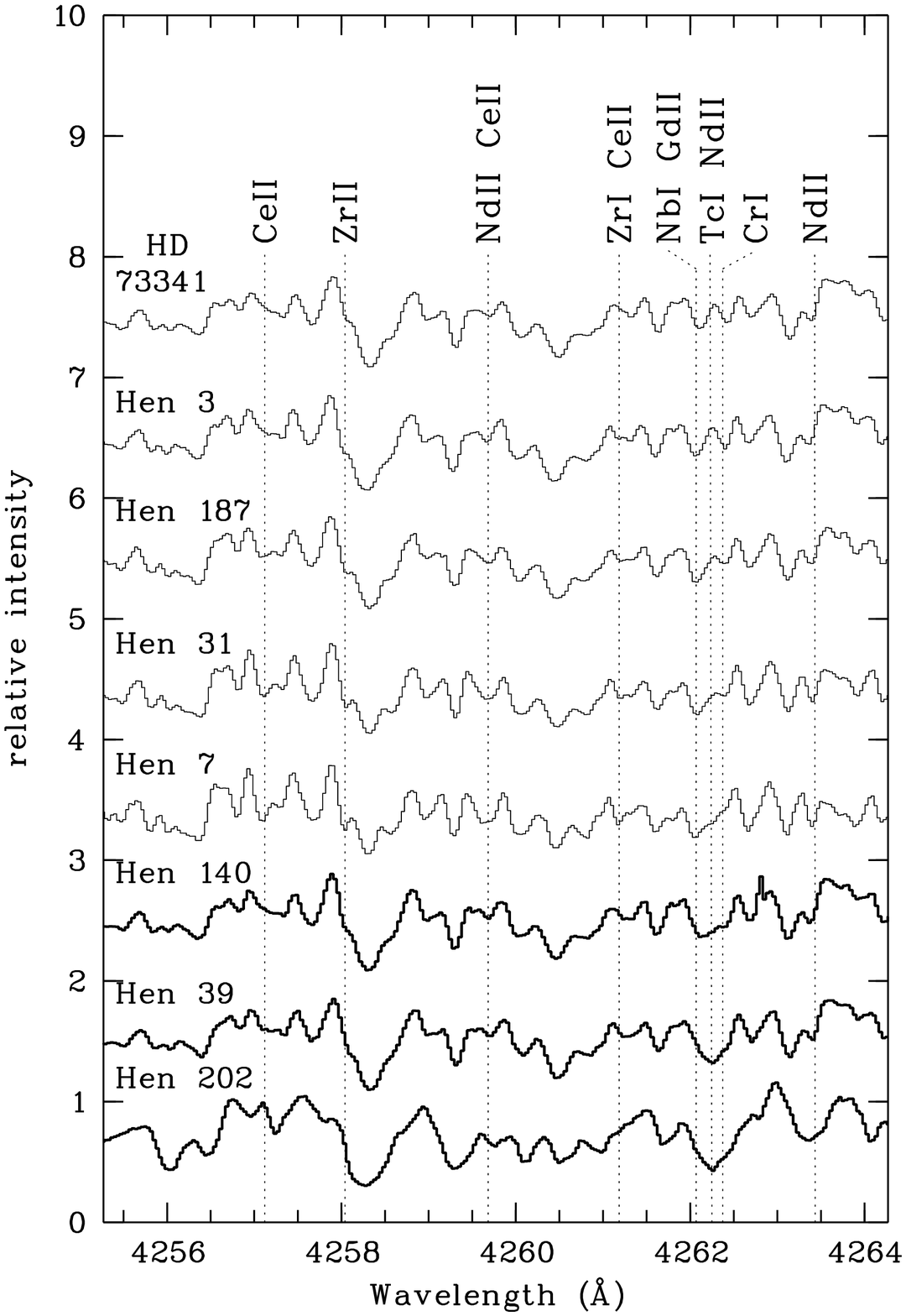,width=8.0cm,height=12.0cm}}
   \end{center}
   \caption{\label{Fig:figTc62}
   Spectra in the 4262\AA~ region. 
   The top spectrum is HD~73341, a normal giant (M3/M4III);
   the other spectra correspond to Henize S stars.
   Technetium-poor spectra are plotted with a thin line; 
   technetium-rich spectra with a thick line.
   Hen 3, 187, 31, 7 are typical Tc-poor stars; Hen 39 and 202
   are typical Tc-rich stars. The Tc-rich spectrum of Hen 140 
   is unique in our sample and remarkable 
   because of its very weak Tc lines (see Sect.~\ref{Sect:Analysis}). 
   All spectra are plotted on the same relative intensity scale.
   The local pseudo-continuum point has been taken as an average
   of the fluxes at 4239.1, 4244.1, 4247.1 and 4265.4\AA;
   for the sake of clarity, each spectrum (except the lowest one) is
   vertically shifted by 1 unit with respect to the spectrum below it.
   Some spectral features of s-process elements are identified 
   (see text)
   }
\end{figure}

\begin{figure}
   \begin{center}
   \leavevmode
   \centerline{\psfig{file=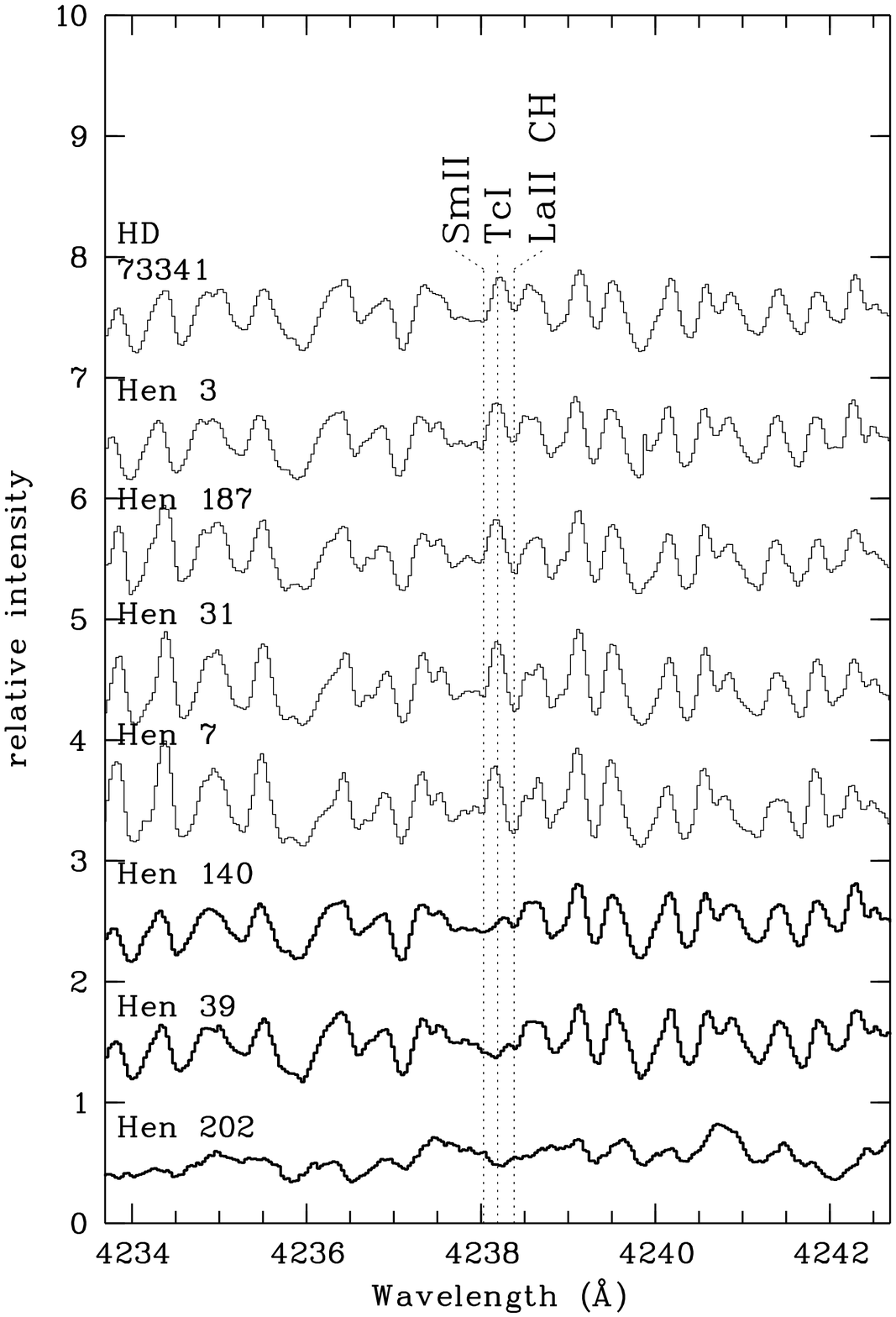,width=8.0cm,height=12.0cm}}
   \end{center}
   \caption{\label{Fig:figTc38}
   Same as Fig.~\ref{Fig:figTc62} for the 4238\AA~ region
   }
\end{figure}

The three strong resonance lines of \ion{Tc}{i} are located 
at 4238\AA, 4262\AA~ and 4297\AA, with intensity
ratios of 3:4:5. All three lines are severely blended (Little-Marenin \& 
Little 1979, their Table III).
With the adopted instrumental configurations, a single exposure 
spans 35 to 50\AA; it is thus possible to observe 
simultaneously the 4238\AA~ and 4262\AA~ lines.
In this analysis we follow the guidelines provided in the
landmark paper of Smith \& Lambert (\cite{Smith})
and therefore concentrate on the most useful 4262\AA~ line, while the
4238\AA~ line is used as an independent confirmation.

Fig.~\ref{Fig:figTc62} shows ex\-am\-ples of spec\-tra in the 4262\AA~ 
re\-gion for an M3-4 giant (HD~73341)
and for seven S stars (four being technetium-poor: Hen 3, 187, 31, 7
and three technetium-rich: Hen 140, 39 and 202 = $\pi^1$ Gru).
It can be seen that the Tc $\lambda 4262.270$\AA~ \- line is blended with
two features; the bluest includes primarily
\ion{Nb}{i} (4262.050\AA) and \ion{Gd}{ii} (4262.087\AA), 
and the reddest \ion{Cr}{i} (4262.373\AA) 
(see Fig.~\ref{Fig:figTc62}).
A weaker contribution of \ion{Nd}{ii} at 4262.228\AA,
almost on the top of the \ion{Tc}{i} line, may also be present.
These composite features are much weaker than the \ion{Tc}{i}
resonance line at its maximum strength; moreover,
the \ion{Nb}{i}-\ion{Gd}{ii} blend and the \ion{Tc}{i} line are 0.18\AA~ apart.
Therefore the shape and location of the 
\ion{Nb}{i}-\ion{Gd}{ii} (-\ion{Tc}{i}) 
blend (hereafter called $X_{4262}$ feature)
clearly depends on whether it contains the technetium line or not.
Quantitatively, the minimum of a gaussian fitted to a Tc-containing
$X_{4262}$ feature is shifted redward by $\sim0.14$\AA~
with respect to the minimum of a gaussian fitted to a no-Tc $X_{4262}$ feature;
such a shift is easily detectable on our spectra (compare Hen 3 or Hen 7
with Hen 39 on Fig.~\ref{Fig:figTc62}).

In practice, each spectrum has been rebinned to zero-redshift
in the following way:
10 nearby ($\le 5$\AA~ on either side) apparently unblended 
stellar features with unambiguous identification, 
are adopted as wavelength standards.
Gaussian profiles are fitted to these lines and provide
a mean redshift.
The wavelength of the $X_{4262}$ feature is then computed as the
minimum of a gaussian centered on the $X_{4262}$ feature
of the redshift-corrected spectrum.
Typical uncertainties on the $X_{4262}$ wavelength
amount to 0.013\AA~ for technetium-poor stars and 
0.020\AA~ for technetium-rich stars 
(as derived from the standard deviation on the mean redshift).

The same method is applied to the 4238\AA~ technetium line,
where the CH-\ion{La}{ii} blend has been taken as the $X_{4238}$ feature
(Fig.~\ref{Fig:figTc38}). 
Typical uncertainties on the $X_{4238}$ wavelength are slightly
larger (0.023\AA~ for Tc-rich S stars and 0.025\AA~ for Tc-poor S stars)
because of the lower S/N ratio and the stronger blending at 4238\AA.
Results are listed in Tables~\ref{Tab1} and ~\ref{Tab2};
the $X_{4238}$ and $X_{4262}$ features always yield consistent results
regarding the absence or presence of technetium, except for Hen 140 
(=HD 120179).

This star is indeed unique in having very weak technetium features
(see Figs.~\ref{Fig:figTc62} and \ref{Fig:figTc38}). A second spectrum,
taken 3.5 years later, is almost identical to the one displayed 
in  Fig.~\ref{Fig:figTc62} and \ref{Fig:figTc38}.
The blind application of gaussian fitting to the $X_{4238}$ feature
of Hen 140 yields a central wavelength that would qualify it as Tc-poor;
however, the extreme weakness of the pseudo-emission separating the
LaII-CH blend from the SmII line 
as seen in Hen 140 (Fig.~\ref{Fig:figTc38})
is unusual for Tc-poor stars, and suggests the presence of a weak
technetium line, as confirmed from the appearance of the $X_{4262}$
feature. We therefore believe that Hen 140 is the unique example
in our sample of an S star with very weak Tc lines.

In all the other cases, technetium (non-) detection
relies on the location of the minimum of a gaussian fitted
to the $X_{4262}$ (or $X_{4238}$) blend.

\medskip

Are there, with this method, risks 
(i) to misclassify as Tc-rich a truly Tc-poor star, and
(ii) to misclassify as Tc-poor a truly Tc-rich star?
We show in the remaining of this section that both risks are
most probably non-existent in the present study.

\begin{sloppypar}
Error (i) could, in principle, affect very luminous Tc-poor stars, 
because their large macroturbulence would   
broaden their $X_{4262}$ feature, 
which could then possibly mimick a Tc-rich feature.
\end{sloppypar}

In order to test this hypothesis, gaussian filters of different widths
have been applied to Tc-poor spectra, so as to make their line widths
comparable to those of the stars classified as Tc-rich.
This simulation clearly shows that even the largest macroturbulence
value observed in our sample (T Ceti) is not large enough to make
truly Tc-poor stars appear as Tc-rich from the broadening of their
$X_{4262}$ feature. That conclusion is even more stringent 
when considering the $X_{4238}$ feature.
However, this risk cannot be excluded for very luminous stars
(class I or II) if observed at lower resolution (R $< 30\,000$).

Error (ii) could, in principle, occur for stars displaying 
a `weak technetium line'
(weaker than the Tc lines of Hen 140 discussed above)
with an intensity not large enough to shift 
the $X_{4262}$ blend redward from the Tc-poor wavelength.
In fact, some stars in our sample exhibit an `ambiguous'
$X_{4262}$ blend, in the sense that the pseudo-emission 
located between the \ion{Nb}{i}-\ion{Gd}{ii} lines 
and the \ion{Cr}{i} line becomes very weak or even disappears,
mimicking a `weak technetium line' (a typical example is Hen 7
on Fig.~\ref{Fig:figTc62}).
Such a star is classified by our method as technetium-poor,
for the minimum of the $X_{4262}$ blend remains unchanged with
respect to the no-Tc cases.

In fact, all intermediates exist between the `unambiguous' 
$X_{4262}$ Tc-poor blends (with a clear central pseudo-emission,
see HD~73341 and Hen 3 on Fig.~\ref{Fig:figTc62}) 
and the `ambiguous' $X_{4262}$ blends (where this pseudo-emission 
is absent, as in Hen 7); 
two typical transition cases are plotted on Fig.~\ref{Fig:figTc62} 
(Hen 187 and Hen 31).
These `ambiguous' spectra were taken during different observing
runs; the shape of the $X_{4262}$ blend is independent of
the resolution and of the S/N ratio of the spectra.

Do these `ambiguous' spectra correspond to stars with a weak technetium
line, intermediate between the clear Tc-poor and Tc-rich cases?
In fact these `ambiguous' spectra are clearly different from
the spectum of the weakly Tc-rich star Hen 140, 
for {\it their $X_{4238}$ feature is identical to the $X_{4238}$ feature 
of the unambiguous Tc-poor stars} (Fig.~\ref{Fig:figTc38}),
which clearly indicates that technetium is absent in these stars.

It may therefore be concluded that our method of gaussian fit
to the $X_{4238}$ and $X_{4262}$ features is able to properly
separate technetium-rich from technetium-poor S stars.

What then causes the variety of $X_{4262}$ features observed
in Fig.~\ref{Fig:figTc62} for technetium-poor S stars?
The spectral sequence going from Hen 3 to Hen 7 on Fig.~\ref{Fig:figTc62}
is not a temperature sequence. The temperature of the
stars of our sample have been derived from the $V-K$ color index 
using the Ridgway et al. (\cite{Ridgway}) calibration, 
the $K$ magnitudes from Catchpole et al. (\cite{Catchpole79})
and our Geneva photometry.
Although the bulk of technetium-rich S stars are clearly cooler than
technetium-poor S stars (see also Van Eck et al. \cite{Van Eck98}),
there is no sign whatsoever of a possible correlation between
the shape of the $X_{4262}$ blend of technetium-poor stars and their 
temperature.
MOOG (Sneden \cite {Sneden}) synthetic spectra 
(for stars with T$_{eff}\sim 3400-3800$K
as derived from their $V-K$ index) indicate
that neither gravity nor metallicity 
can significantly modify the $X_{4262}$ blend.

A closer inspection of the spectral sequence of 
Fig.~\ref{Fig:figTc62} (from HD~73341 to Hen 7)
reveals that several lines become stronger as the central 
pseudo-emission of the $X_{4262}$ blend weakens. 
The major contributors to these features, identified with the help of
synthetic spectra, are indicated on the top of Fig.~\ref{Fig:figTc62}.
It is noteworthy that {\it all these elements are s-process elements}.
The sequence of spectra (drawn with a thin line) in Fig.~\ref{Fig:figTc62} 
is thus, from
top to bottom, a sequence of increasing s-process line strengths
(s-process lines being weak, as expected, in the M star
HD~73341).
The line which progressively blends the $X_{4262}$ feature of 
technetium-poor stars
is thus probably an s-process line as well. Since it cannot be technetium
(see above), a good candidate
is the 4262.228\AA~ line of \ion{Nd}{ii}, or perhaps 
the wing of the \ion{Gd}{ii} line at 4262.087\AA.

It is not surprising to find a wide range of s-process enhancements
among technetium-poor S stars, since these stars have accreted
their s-process-enriched matter from a companion star.
Hence the level of chemical peculiarities is not linked
to the evolutionary status of the star, but rather depends upon
the amount of s-process accreted matter (see Jorissen et al. 
\cite{Jorissen} for a detailed discussion).

These s-process lines are more difficult to see in the technetium-rich
S stars, probably because in these cooler and more luminous stars, 
lines are broader 
(because of a larger macroturbulence) and the molecular 
blanketing is more severe.

\subsection{Misclassified and SC stars}
\label{Sect:SC}

\begin{figure}
   \begin{center}
   \leavevmode
   \centerline{\psfig{file=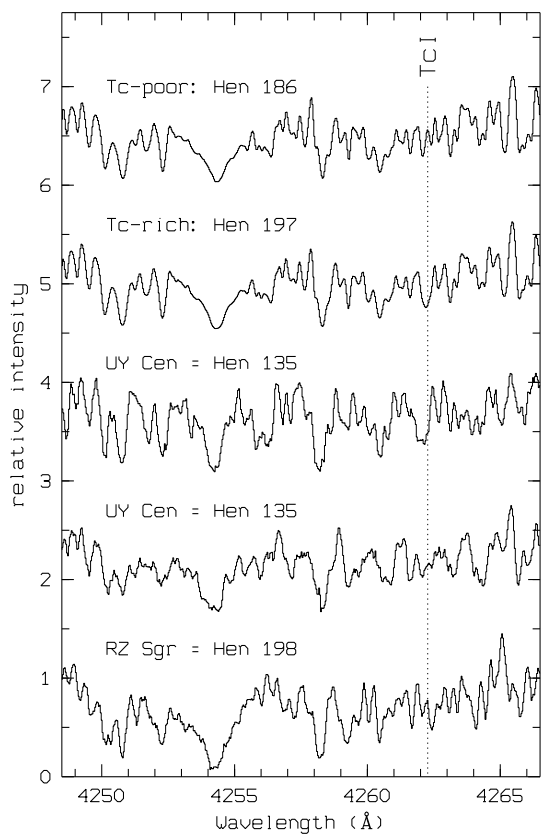,width=8.0cm,height=12.0cm}}
   \end{center}
   \caption{\label{Fig:figTc62except}
   Same as Fig.~\ref{Fig:figTc62}, but for two prototype S stars 
   (Hen 186, Tc-poor and Hen 197, Tc-rich),
   and for the two outstanding stars UY Cen (SC) and RZ Sgr.
   Note the striking differences between the lower and upper spectra 
   of UY Cen, taken on March 16, 1993
   and on January 31, 1997, respectively.
   The local pseudo-continuum point is defined 
   as in Fig.~\ref{Fig:figTc62};
   for the sake of clarity, each spectrum (except the lowest one) is
   vertically shifted by 1.5 unit with respect to the spectrum below it
   }
   
\end{figure}

The method outlined in Sect.~\ref{Sect:Analysis} cannot be applied to
four stars of our sample which exhibit peculiar spectra 
(Hen 22, 135, 154 and 198).
In order to check the assignment of the Henize stars to spectral type S,
low-resolution spectra
($\Delta \lambda \sim 0.3$ nm, $4400$\AA $<\lambda< 8200$\AA) 
have been obtained for all stars from Hen 3 to Hen 165
at ESO on the 1.52m telescope equipped with the Boller \& Chivens
spectrograph and grating \#23 (Van Eck et al. \cite{Van Eck}).

Two misclassified stars have been uncovered: Hen 22 and Hen 154
show no sign of ZrO bands whatsoever in their spectra.
Besides, Hen 22 is classified as `S:' by Henize.
Both stars cannot be dwarfs because their NaD and MgH $\lambda$4780\AA~
features are too weak. 
Their prominent \ion{Ca}{i} $\lambda$ 4455\AA~ line and
their weak CN $\lambda$7895\AA~ band point towards them being giant stars
rather than supergiants. 
Type Ia supergiants can certainly be ruled out because
their absolute magnitudes 
($M_{\rm v}=-7.8$ and -7.5 for G8Ia and K3-5I respectively, 
Landolt-B\"ornstein \cite{Landolt}) would result in
much too large heights above the galactic plane 
(7.5 kpc for Hen 154 and 13.2 kpc for Hen 22).
Hen 154 is probably a late G giant ($\sim$G8), and Hen 22 
a mid-K giant ($\sim$K3-5). These assignments are compatible 
with the Geneva photometry available for these two stars.

As far as Hen 135 and Hen 198 are concerned, 
Fig.~\ref{Fig:figTc62except} shows that the spectra of these stars 
are very different from those of other S stars 
of the Henize sample.
Many spectral features adopted as wavelength standards,
as well as the technetium blend, are difficult or even impossible 
to identify in the spectra of Hen 135 and Hen 198.
In fact, we show below that these two stars are the only
two SC stars in the subsample of 
Henize S stars observed with the CAT\footnote{
  Although the Henize sample contains several SC stars,
  they are usually too red, hence too faint at 4250\AA,
  to be observed with the CAT.}.
Hen 135 ($V\sim 7$) and Hen 198 ($V\sim 7-10$) were indeed 
the only very red stars ($B-V>2$)
which were bright enough to allow spectra to be taken
in the violet.

SC stars are known to have very peculiar spectra. Their spectrum is
filled with strong atomic lines and almost no molecular bands
in the optical, a consequence of their C/O ratio being very close to unity
(Scalo \cite{Scalo}). 
Catchpole \& Feast (\cite{Catchpole71}) define SC stars 
from the following three criteria:
(i) extremely strong Na D lines, 
(ii) drop in the continuum intensity shortward of 4500\AA, and 
(iii) bands of ZrO and CN simultaneously present (though quite weak),
as well as general resemblance of the spectrum 
(i.e. regarding `the absolute and relative
strength of metal lines') with that of UY Cen.

Hen 135 (=UY Cen) is thus the prototype SC star.
Our two spectra of that star (taken in March 1993 and January 1997,
see Fig.~\ref{Fig:figTc62except})
are quite different;
in particular the shape of the $X_{4262}$ feature has changed noticeably.
Therefore it is hazardous to infer the technetium content of UY Cen 
from these data alone without the help of appropriate model atmospheres
and synthetic spectra, which is beyond the scope of this paper.

Hen 198 (=RZ Sgr) has an Se-type spectrum;
Stephenson (\cite{Stephenson}) quotes the HD catalogue noting that 
`the spectrum is similar to class N, but does not belong to that class'. 
It is probably associated with a reflection nebula 
(Whitelock \cite{Whitelock94}).
RZ Sgr is a large-amplitude ($\sim2.5$ mag) SRb-type variable ($P=203.6$ d).
Its H$\alpha$ emission, as well as the TiO and ZrO band strengths, 
are variable.
Catchpole \& Feast (\cite{Catchpole76}) also note that the Zr:Ti 
ratio of RZ Sgr
is unusually high for an S star, and rather close to the one of
N-type carbon stars. 

Although RZ Sgr has not been classified as an SC star, 
it shares many common features with that family.
Indeed, it reasonably meets the three criteria mentioned
above for SC stars:\\
\noindent(i) Reid and Mould (\cite{Reid}) measured 
the strength of the Na D lines
for several S, SC and C stars, including RZ Sgr. 
A spectrophotometric index of 1.07 is found for RZ Sgr, much larger 
than typical values for
S stars (0.22 for \BD{+28}{4592} and 0.28 for NQ Pup),
but comparable to values obtained for SC stars 
(0.55 for LMC 441, 1.73 for R CMi, 2.57 for VX Aql).
Thus RZ Sgr has abnormally strong Na D lines with respect to other S stars.

\noindent(ii) The ultraviolet flux deficiency of SC stars is clearly 
apparent from photometric data in the Geneva system. 
Indeed, the mean wavelengths of the $B$ and $V$ filters
are $\lambda_0(B)$ = 4227\AA~ and $\lambda_0(V)$ = 5488\AA~
(Rufener \& Nicolet \cite{Rufener});
therefore the $B-V$ index is highly sensitive to the ultraviolet
flux deficiency of SC stars occurring for $\lambda < 4500$\AA.
SC stars have $B-V>2$, whereas the bulk of S stars have $B-V<2$. 
In that respect again, RZ Sgr ($2.0<B-V<3.0$) is typical of SC stars.

\noindent(iii) ZrO is present (although weak) in RZ Sgr; we found no
information about the possible presence of CN bands. 
Infrared CO bands are stronger in RZ Sgr than in many other S and SC stars 
(Whitelock et al. \cite{Whitelock85}), probably locking a great quantity 
of carbon.

The IRAS colours of RZ Sgr also share many similarities with SC stars:
it is located in a region of the ($K-[12]$,$[25]-[60]$) color-color diagram
(`region E' as defined by Jorissen \& Knapp \cite{JorissenKnapp}) 
containing mainly SC stars with large 60$\mu$m excess 
and often resolved shells (see also Young et al. \cite{Young}).

All these arguments therefore indicate that RZ Sgr is closely 
related to the SC family. As pointed out for UY Cen, 
the 4262\AA~ and 4238\AA~ lines of \ion{Tc}{i}
are very difficult to analyse in SC stars.
An assignment of these two stars to either the Tc-rich or Tc-poor
group has therefore not been attempted here. Abia \& Wallerstein (\cite{Abia})
nevertheless suggest that SC stars are Tc-rich, based on a 
quantitative analysis.

\section{Discussion}
\label{Sect:RESULTS}

\subsection{The technetium dichotomy}

\begin{figure}
   \begin{center}
   \leavevmode
   \centerline{\psfig{file=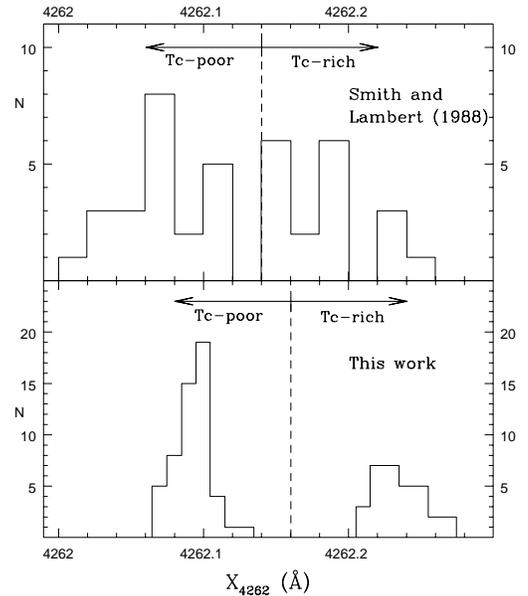,width=8.0cm,height=8.0cm}}
   \end{center}
   \caption{\label{Fig:Tc}
   Frequency histograms of the wavelength of the $X_{4262}$ spectral 
   feature. 
   Top: results of Smith \& Lambert (\cite{Smith}) for their sample of 
MS and S stars; 
   the dotted line delimits the boundary wavelength (4262.14\AA)
   between Tc-poor and Tc-rich S stars, as adopted by Smith \& Lambert 
(\cite{Smith}).
   Bottom: same for all stars of Tables~\ref{Tab1} and ~\ref{Tab2}, 
where the boundary wavelength has
   been taken at 4262.16\AA
   }
\end{figure}

\begin{figure}
   \begin{center}
   \leavevmode
   \centerline{\psfig{file=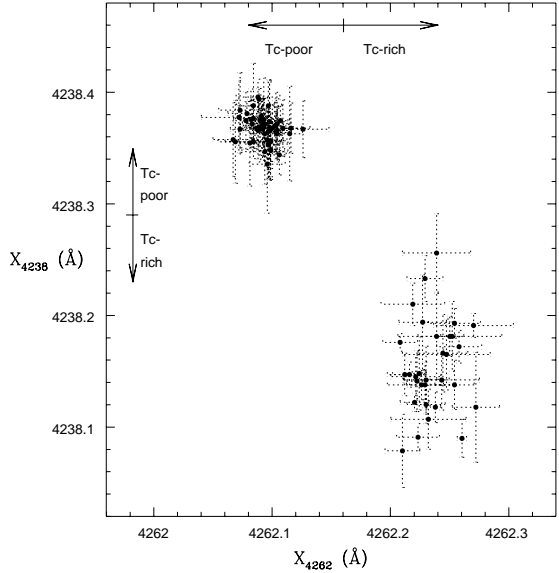,width=8.0cm,height=8.0cm}}
   \end{center}
   \caption{\label{Fig:Tc3862}
     The wavelength of the $X_{4262}$ Tc feature versus 
     the wavelength of the $X_{4238}$ Tc feature. 
     Errorbars represent the standard deviation of the mean redshift
     (computed from the $~\sim$10 spectral features adopted as wavelength 
standards).
     The technetium class (Tc-rich or Tc-poor) derived from 
     the $X_{4238}$ blend always confirms the technetium class
     derived from the $X_{4262}$ blend; the Tc-poor and Tc-rich regions
     are clearly distinct.
     The boundary wavelengths adopted in this study 
     between Tc-rich and Tc-poor stars are
     4262.16\AA~ for the $X_{4262}$ feature (horizontal axis) and
     4238.29\AA~ for the $X_{4238}$ feature (vertical axis);
     they are indicated by arrows
     }
\end{figure}

The lower part of Fig.~\ref{Fig:Tc} shows the 
frequency histogram of the wavelength of the $X_{4262}$ spectral 
feature for stars of Tables~\ref{Tab1} and ~\ref{Tab2}.
The stars of our sample clearly segregate in two groups.
The average wavelength of the bluer group is 4262.093\AA;
this group thus corresponds to Tc-poor S stars. 
The average wavelength of the redder group is 4262.235\AA,
thus revealing the contribution of the \ion{Tc}{i} 4262.270\AA~ line
to the \ion{Gd}{ii}-\ion{Nb}{i} blend.
The standard deviation on the $X_{4262}$ wavelengths is
0.012\AA~ for Tc-poor stars and 0.017\AA~ for Tc-rich stars.
These values are in good agreement with 
the estimated errors on the $X_{4262}$ wavelength 
(0.013\AA~ for Tc-poor and 0.020\AA~ for Tc-rich stars, \
Sect.~\ref{Sect:Analysis}).

The two groups are clearly separated by a 0.08\AA~ gap, 
{\it with no intermediate cases}. Therefore, in order
to distinguish Tc-poor from Tc-rich stars (on our spectra
of resolution in the range 30\,000-60\,000),
a delimiting wavelength of 4262.16\AA~ may be safely adopted.

A similar conclusion holds for the
$X_{4238}$ feature (Fig.~\ref{Fig:Tc3862}), where a boundary wavelength
of 4238.29\AA~ unambiguously separates the two kinds of S stars.
Fig.~\ref{Fig:Tc3862} further shows that the diagnostics provided
by the $X_{4238}$ and $X_{4262}$ features are consistent
with each other.

For comparison purpose the frequency histogram of the wavelength 
of the $X_{4262}$ spectral feature as obtained by 
Smith \& Lambert (\cite{Smith})
is plotted in the upper part of Fig.~\ref{Fig:Tc}, for their sample
of MS and S stars (their Table 2).
The segregation into Tc-poor and Tc-rich S stars (with 4 stars falling
on their boundary wavelength at 4262.14\AA)
is not as clean as with our higher resolution spectra.

The small number of `transition stars' in our sample 
(i.e. stars with weak Tc lines, the only case being Hen 140)
is noteworthy. This result may provide constraints on the evolution
with time of the technetium abundance along the TPAGB 
(Smith \& Lambert \cite{Smith}; Busso et al. \cite{Busso})
and clearly deserves further studies. For example, it would be 
of interest to investigate whether the small number of S stars
with weak technetium lines found in our sample implies 
that the very first objects to dredge-up heavy elements 
on the TPAGB are not S stars but rather M stars.
Indeed, the Tc detection threshold might not coincide with the ZrO
detection threshold, but be slightly lower 
(i.e. {\it stars would appear Tc-rich before being ZrO-rich}); 
this would explain the puzzling 
Tc-rich M stars discovered by Little-Marenin \& Little (\cite{Little-Marenin}) 
and Little et al. (\cite{Little}).
It is moreover necessary to disentangle
abundance effects from atmospheric effects on the technetium line
strength. 

Although a more detailed study is deferred to a forthcoming 
paper, it may already be mentioned at this point that the
Tc/no-Tc dichotomy reported in this paper is not due to technetium
being entirely ionized in the warmer S stars. Indeed, 
the \ion{Tc}{i}/(\ion{Tc}{i} + \ion{Tc}{ii}) ratio is still 
$\sim40\%$ in the warmest S stars (T$_{eff}$=3800 K) 
while it amounts to $\sim70\%$ at T$_{eff}$=3500 K 
and to $\sim95\%$ at T$_{eff}$=3000~K, according to the Saha ionization
equilibrium formula (with representative electron densities 
taken from model atmospheres).

\subsection{M stars with 60$\mu$m excess and symbiotic stars}

None of the four M stars with 60$\mu$m excess taken from the sample 
of Zijlstra et al. (\cite {Zijlstra}) show technetium.
This observation clearly indicates that these stars,
which are surrounded by cool dust dating back to a former episode of
strong mass loss, do not currently experience heavy elements synthesis
followed by third dredge-ups.
The same conclusion holds true for the two observed symbiotic stars
(SY Mus and RW Hya).

\section{Conclusion}
High-resolution spectra have
been obtained and analysed to infer the technetium content
of 76 S, 8 M and 2 symbiotic stars.
The presence or absence of technetium was deduced from
the shape of two blends involving technetium at 4238\AA~ and 4262\AA~  
(more precisely: from the wavelength of their minimum).
However this method does not apply to SC stars. 
Two misclassified S stars (Hen 22 and Hen 154) have emerged.
The technetium (non-)detection at 4238\AA~ is consistent with
the result at 4262\AA. 
Only one `transition' case (Hen 140 = HD 120179, a star where
only weak lines of technetium are detectable) is found in
our sample. 

A resolution in excess of $30\,000$
is definitely required to provide unambiguous conclusions
regarding presence or absence of technetium. For example,
at 4262\AA, an s-process line (possibly \ion{Nd}{ii}) 
is suspected to sometimes mimick a weak technetium line
(although the 4238\AA~ feature clearly shows that technetium
is absent).
The shape of the $\lambda$4262\AA~ feature varies from one
Tc-poor star to another, depending on the s-process overabundance
level, which is in turn a function of the amount of accreted matter
by these binary S stars.

Among the 70 analysed Henize S stars, 41 turn out to be technetium-poor
and 29 tech\-ne\-tium-rich. That fraction may not be used, however,
to infer the relative frequencies of intrinsic and extrinsic S stars,
since the subsample of Henize S stars observed with the CAT is 
biased towards the brightest and bluest stars.
The frequency of extrinsic/intrinsic S stars will be derived from
the whole data set in a forthcoming paper.

\acknowledgements{
This research has made use of the Simbad database,
operated at CDS, Strasbourg, France. S.V.E. thanks F.R.I.A. (Belgium) 
for financial support.}


\begin{thebibliography}{}
\bibitem[1998]{Abia}
  Abia C., Wallerstein G., 1998, MNRAS 293, 89
\bibitem[1990]{Brown}
  Brown J.A., Smith V.V., Lambert D.L., Dutchover E.Jr., Hinkle K.H.,
  Johnson H.R., 1990, AJ 99, 1930
\bibitem[1992]{Busso}
  Busso M., Gallino R., Lambert D.L., Raiteri C.M., Smith V.V., 1992, 
ApJ 399, 218
\bibitem[1971]{Catchpole71}
  Catchpole R.M., Feast M.W., 1971, MNRAS 154, 197
\bibitem[1976]{Catchpole76}
  Catchpole R.M., Feast M.W., 1976, MNRAS 175, 501
\bibitem[1979]{Catchpole79}
  Catchpole R.M., Robertson B.S.C., Lloyd Evans T.H.H., Feast M.W., 
  Glass I.S., Carter B.S., 1979, Circulars of the South African Astronomical
  Observatory, Vol 1, Nr 4, p 61
\bibitem[1995]{Chen}
  Chen P.S., Gao H., Jorissen A., 1995, A\&AS 113, 51
\bibitem[1984]{Cosner}
  Cosner K.R., Despain K.H., Truran J.W., 1984, ApJ 283, 313
\bibitem[1986]{Horne}
  Horne K., 1986, PASP 98, 609
\bibitem[1993]{Johnson}
  Johnson H.R., Ake T.B., Ameen M.M., 1993, ApJ 402, 667
\bibitem[1993]{Jorissen93}
  Jorissen A., Frayer D.T., Johnson H.R., Mayor M., Smith V.V., 1993,
  A\&A 271, 463
\bibitem[1998]{JorissenKnapp}
  Jorissen A., Knapp G. R., 1998, A\&AS 129, 363
\bibitem[1998]{Jorissen}
  Jorissen A., Van Eck S., Mayor M., Udry S., 1998, A\&A 332, 877
\bibitem[1960]{Henize}
  Henize K.G., 1960, AJ 65, 491
\bibitem[1996]{Kaper}
  Kaper L., Pasquini L., 1996, CAT+CES Operating Manual, ESO, 
3p6CAT-MAN-0633-0001
\bibitem[1982]{Landolt}
  Landolt-B\"ornstein, Numerical Data and Functional Relationships in Science
  and Technology, New Series, Hellwege ed., Astronomy and Astrophysics, 
  vol 2, subvol b, 1982
\bibitem[1989]{Lindgren}
  Lindgren H., Gilliotte A., 1989, The Coud\'e Echelle Spectrometer, 
The Coud\'e Auxiliary Telescope, ESO Operating Manual No. 8
\bibitem[1979]{Little-Marenin}
  Little-Marenin I.R., Little S.J., 1979, AJ 84, 1374 
\bibitem[1987]{Little}
  Little S.J., Little-Marenin I.R., Hagen-Bauer W., 1987, AJ94, 981
\bibitem[1982]{MacConnell}
  MacConnell D.J., 1982, A\&AS 48, 355
\bibitem[1986]{Mathews}
  Mathews G.J., Takahashi K., Ward R.A., Howard W.M., 1986, ApJ 302, 410
\bibitem[1922]{Merrill22}
  Merrill P.W., 1922, ApJ 56, 457
\bibitem[1952]{Merrill52}
  Merrill P.W., 1952, ApJ 116, 21
\bibitem[1997]{Mowlavi}
  Mowlavi N., 1997, in: Tours Symposium on Nuclear Physics III, 
   eds. M. Arnould, M. Lewitowicz, Y.T. Oganessian, M. Ohta, 
   H. Utsunomiya, T. Wada, AIP Conf. Proc. 425, p507
\bibitem[1991]{Newberry}
  Newberry M.V., 1991, PASP 103, 122
\bibitem[1985]{Reid}
  Reid N., Mould J., 1985, ApJ, 299,236
\bibitem[1980]{Ridgway}
  Ridgway S.T., Joyce R.R., White N.M., Wing R.F., 1980, ApJ 235, 126
\bibitem[1988]{Rufener}
  Rufener F., Nicolet B., 1988, A\&A 206, 357
\bibitem[1973]{Scalo}
  Scalo J.M., 1973, ApJ 186, 967
\bibitem[1974]{Sneden}
  Sneden C., 1974, Ph. D. Thesis, University of Texas, Austin
\bibitem[1988]{Smith}
  Smith V.V., Lambert D.L., 1988, ApJ 333, 219
\bibitem[1990]{Smith90} 
  Smith V.V., Lambert D.L.,  1990, ApJS 72, 387
\bibitem[1984]{Stephenson}
  Stephenson C.B., 1984, The General Catalogue of Galactic S Stars, 
Publ. Warner \& Swasey Observ. 3,1
\bibitem[1995]{Straniero}
  Straniero O., Gallino R., Busso M., Chiefei A., Raiteri C. M., 
Limongi M., Salaris M., 1995, ApJ 440, L85
\bibitem[1998]{Udry}
  Udry S., Mayor M., Van Eck S., Jorissen A., 1998, A\&AS 131, 25
\bibitem[1998]{Van Eck98}
  Van Eck S., Jorissen A., Udry S., Mayor M., Pernier B., 1998, A\&A 329, 971
\bibitem[1999]{Van Eck}
  Van Eck S., Jorissen A., 1999, in preparation
\bibitem[1994]{Whitelock94}
  Whitelock P., 1994, MNRAS 270, L15
\bibitem[1985]{Whitelock85}
  Whitelock P., Catchpole R.M., 1985, MNRAS 212, 873
\bibitem[1993]{Young}
  Young K., Phillips T.G., Knapp G.R., 1993, ApJS 86, 517
\bibitem[1992]{Zijlstra}
  Zijlstra A.A., Loup C., Waters L.B.F.M., de Jong T., 1992, A\&A 265, L5

\end{thebibliography}
\end{document}